\documentclass{article}

\PassOptionsToPackage{numbers,compress,sort}{natbib}

\usepackage[final,main]{neurips_2025}

\usepackage[utf8]{inputenc} %
\usepackage[T1]{fontenc}    %
\usepackage[hyphens]{url}   %
\usepackage{hyperref}       %
\usepackage{booktabs}       %
\usepackage{amsfonts}       %
\usepackage{nicefrac}       %
\usepackage{microtype}      %
\usepackage{xcolor}         %

\usepackage{paralist}
\usepackage{pifont}

\usepackage{graphicx}
	\setkeys{Gin}{width=\textwidth, height=\textheight, keepaspectratio}
\usepackage[skip=0pt]{subcaption}

\newcommand{\dpmm}[1]{\emph{dpmm}}
\newcommand{\descr}[1]{\noindent\textbf{#1}}

\title{\dpmm{}: Differentially Private Marginal Models,\\a Library for Synthetic Tabular Data Generation}

\author{%
  Sofiane Mahiou$^1$ \quad Amir Dizche$^1$ \quad Reza Nazari$^1$ \quad Xinmin Wu$^1$ \\
  \textbf{Ralph Abbey}$^1$ \quad \textbf{Jorge Silva}$^1$ \quad \textbf{Georgi Ganev}$^{1,2}$ \\
  $^1$SAS \quad $^2$UCL\\
  \texttt{\href{mailto:sofiane.mahiou@sas.com}{sofiane.mahiou@sas.com}}\\
}

\begin{document}

\maketitle

\begin{abstract}
We propose \dpmm{}, an open-source library for synthetic data generation with Differentially Private (DP) guarantees.
It includes three popular marginal models -- PrivBayes, MST, and AIM -- that achieve superior utility and offer richer functionality compared to alternative implementations.
Additionally, we adopt best practices to provide end-to-end DP guarantees and address well-known DP-related vulnerabilities.
Our goal is to accommodate a wide audience with easy-to-install, highly customizable, and robust model implementations.

Our codebase is available from: \url{https://github.com/sassoftware/dpmm}.
\end{abstract}

\section{Introduction}
Privacy-preserving synthetic data, leveraging rigorous privacy definitions like Differential Privacy~(DP), offers a promising solution for sharing sensitive data within and across organizations while safeguarding individual privacy.
Interest in synthetic data has been growing not only in the research community~\citep{jordon2022synthetic, cristofaro2024synthetic, hu2024sok}, but it has also been adopted by government agencies for sharing census data~\citep{nasem2020census, ons2023synthesising, hod2025differentially}, as well as in finance and healthcare sectors~\citep{microsoft2022iom, ico2023synthetic, fca2024using}.
The underlying idea is to train generative machine learning models on sensitive data while satisfying the definition of DP through ingesting well-calibrated randomness and noise~\citep{dwork2006calibrating, dwork2014algorithmic}.

While there are numerous proposed DP generative modeling approaches in the literature~\cite{li2014differentially, zhang2017privbayes, xie2018differentially, zhang2018differentially, jordon2018pate, acs2018differentially, abay2018privacy, chanyaswad2019ron, vietri2020new, cai2021data, long2021gpate, aydore2021differentially, zhang2021privsyn, liu2021iterative, mckenna2021winning, mckenna2022aim, vietri2022private}, marginal models, particularly MST~\citep{mckenna2021winning} and AIM~\citep{mckenna2022aim} (and to an extent PrivBayes~\citep{zhang2017privbayes}), have consistently demonstrated strong privacy-utility tradeoffs~\citep{tao2022benchmarking, ganev2024graphical}.
This is especially true for relatively small datasets with fewer than 32 features.
These models broadly rely on the {\em select-measure-generate} paradigm~\citep{mckenna2021winning, mckenna2022simple}, i.e., they select a collection of marginals of interest, measure them while adding noise, and then generate synthetic data that preserves them.

As a result, these three marginal models -- PrivBayes, MST, and AIM -- have been reimplemented by several popular open-source libraries for DP synthetic data generation~\citep{ping2017datasynthesizer, mckenna2019private, opendp2021smartnoise, mahiou2022dpart, qian2023synthcity}.
However, most libraries are cumbersome to install, have outdated dependencies, do not implement all models, and the ones they do implement often have limited functionality.
Moreover, many of the implementations lack actual end-to-end DP pipelines, as the data domain might be directly extracted from the input data, or data preprocessing might be done in a non-DP manner~\cite{annamalai2024you, ganev2025the}.
Additionally, they may contain DP-related bugs such as fixed random states and floating-point vulnerabilities~\citep{haney2022precision, casacuberta2022widespread, lokna2023group}.

To address these limitations, we propose \emph{D}ifferentially \emph{P}rivate \emph{m}arginal \emph{m}odels, or \dpmm{}, a lightweight open-source library (pip installable; Apache-2.0 license) for synthetic data generation.
We incorporate best practices from various scientific papers and DP libraries to provide robust implementations of the three most widely used marginal models with end-to-end DP guarantees and rich functionality (see Sec.~\ref{sec:over}).
Additionally, we achieve higher utility compared to previous implementations and conduct DP auditing experiments (see Sec.~\ref{sec:emp}).

\begin{figure*}[t!]
  \vspace{-0.5cm}
  \begin{subfigure}{0.99\linewidth}
    \centering
    \includegraphics[width=0.66\linewidth]{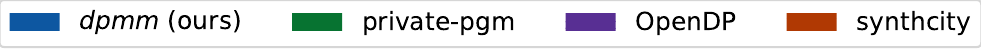}
  \end{subfigure}
	\begin{subfigure}{0.49\linewidth}
    \centering
		\includegraphics[width=0.99\linewidth]{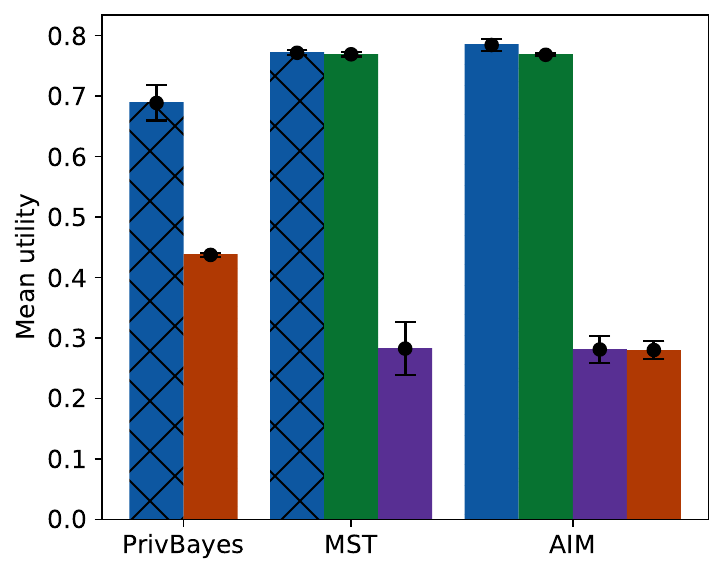}
		\caption{\small Utility}
		\label{fig:utility}
	\end{subfigure}
	\begin{subfigure}{0.49\linewidth}
    \centering
  	\includegraphics[width=0.99\linewidth]{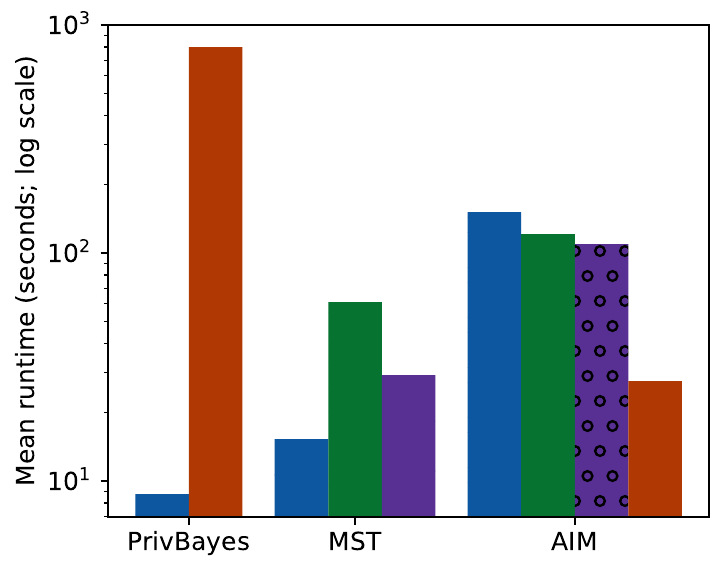}
    \caption{\small Runtime (seconds)}
  	\label{fig:runtime}
	\end{subfigure}
  \caption{\small Comparison between \dpmm{} and other libraries in terms of utility and runtime, all models are trained with default hyperparameters, $\epsilon=1$, and $\delta=10^{-5}$ on Wine.}
  \label{fig:comparisson}
  \vspace{-0.5cm}
\end{figure*}

\descr{Main Contributions:}
\begin{itemize}
  \item We implement and open source \dpmm{}, a lightweight library for end-to-end DP synthetic data generation, containing three popular marginal models with extended functionality -- PrivBayes, MST, and AIM (see left half of Table~\ref{tab:dp} as well as Table~\ref{tab:feat}).
  \item \dpmm{} achieves higher utility than previous implementations -- on average 1.5\% higher than private-pgm and 147\% than OpenDP/synthcity (see Fig.~\ref{fig:utility}) -- due to improved preprocessing.
  \item \dpmm{} contains state-of-the-art DP auditing procedures and effectively addresses known DP-related vulnerabilities (see right half of Table~\ref{tab:dp}).
\end{itemize}

\begin{table}[t!]
  \small
  \centering
  \setlength{\tabcolsep}{4pt}
  \begin{tabular}{l|ccc|ccc}
    \toprule
      \textbf{Library}                                  & \multicolumn{3}{c|}{\textbf{Marginal Model}}                                                                                  & \multicolumn{3}{c}{\textbf{DP Guarantee}}                               \\
                                                        & \textbf{PrivBayes~\citep{zhang2017privbayes}} & \textbf{MST~\citep{mckenna2021winning}} & \textbf{AIM~\citep{mckenna2022aim}} & \textbf{Data}    & \textbf{Data}            & \textbf{Floating-Point}   \\
                                                        &                                               &                                         &                                     & \textbf{Domain}  & \textbf{Preprocessing}   & \textbf{Precision}        \\
    \midrule
      \textbf{\dpmm{} (ours)}                           & \checkmark                                    & \checkmark                              & \checkmark                          & \checkmark       & \checkmark               & \checkmark                \\
      \textbf{private-pgm~\citep{mckenna2019private}}   & \ding{56}                                     & \checkmark                              & \checkmark                          & $\sim^1$         & \ding{56}                & \ding{56}                 \\
      \textbf{OpenDP~\citep{opendp2021smartnoise}}      & \ding{56}                                     & \checkmark                              & \checkmark                          & $\sim^2$         & $\sim^3$                 & \checkmark                \\
      \textbf{synthcity~\citep{qian2023synthcity}}      & \checkmark                                    & \ding{56}                               & \checkmark                          & \ding{56}        & \ding{56}                & \ding{56}                 \\
    \bottomrule
  \end{tabular}
  \vspace{0.1cm}
  \caption{\small Comparison between \dpmm{} and other libraries over marginal models and end-to-end DP support.
  \\$^1$accepts data domain only for discrete data; $^2$data domain cannot be passed as input, extracts data domain in a DP way only for continuous data; $^3$only supports uniform discretization, which is incidentally DP.}
  \label{tab:dp}
  \vspace{-0.3cm}
\end{table}

\begin{table}[t!]
  \small
  \centering
  \setlength{\tabcolsep}{4pt}
  \begin{tabular}{l|cccccc}
    \toprule
      \textbf{Library}                                  & \textbf{Mixed Data}   & \textbf{Conditional}  & \textbf{Public Data}  & \textbf{Structural}   & \textbf{Max}          & \textbf{Serialization}  \\ %
                                                        & \textbf{Support}      & \textbf{Generation}   & \textbf{Pretraining}  & \textbf{Zeros}        & \textbf{Model Size}   &                         \\ %
    \midrule
      \textbf{\dpmm{} (ours)}                           & \checkmark            & \checkmark            & \checkmark            & \checkmark            & \checkmark            & \checkmark              \\ %
      \textbf{private-pgm~\citep{mckenna2019private}}   & \ding{56}             & \ding{56}             & \ding{56}             & MST                   & AIM                   & \ding{56}               \\ %
      \textbf{OpenDP~\citep{opendp2021smartnoise}}      & \checkmark            & \ding{56}             & \ding{56}             & MST                   & AIM                   & \ding{56}               \\ %
      \textbf{synthcity~\citep{qian2023synthcity}}      & \checkmark            & \ding{56}             & \ding{56}             & \ding{56}             & AIM                   & \checkmark              \\ %
    \bottomrule
  \end{tabular}
  \vspace{0.1cm}
  \caption{\small Comparison between \dpmm{} and other libraries over different functionality support.}
  \label{tab:feat}
  \vspace{-0.3cm}
\end{table}

\section{Overview of \dpmm{}}
\label{sec:over}
In this section, we provide an overview of \dpmm{} in terms of the implemented models, the building blocks we use to guarantee end-to-end DP pipeline, and the supported functionality across all models.

\descr{Marginal Models.}
As mentioned, we implement three marginal models in \dpmm{} -- PrivBayes~\citep{zhang2017privbayes}, MST~\citep{mckenna2021winning}, and AIM~\citep{mckenna2022aim}.
We do so because they: i) are known to perform well in the research community~\citep{tao2022benchmarking, ganev2024graphical}, ii) ranked among the best solutions in the NIST DP synthetic data challenge~\citep{nist2018differential}, iii) have been used to release census data by government agencies~\citep{ons2023synthesising, hod2025differentially}, and iv) have been implemented in popular open-source libraries but have known DP-related vulnerabilities~\citep{casacuberta2022widespread, haney2022precision, lokna2023group, annamalai2024you}.

The three models broadly rely on the select-measure-generate paradigm~\citep{mckenna2021winning, mckenna2022simple}.
They primarily differ in the first step, i.e., selecting a collection of marginals of interest.
In the following step they measure these marginals noisily with the Gaussian mechanism~\citep{mcsherry2007mechanism} and then, one could generate synthetic datasets that are consistent with these noisy marginals through Private-PGM~\citep{mckenna2019graphical}.
Next, we discuss how the three models select marginals:

\begin{compactitem}
  \item \textbf{PrivBayes} constructs an optimal Bayesian network of degree $k$ by optimizing the mutual information scores between the nodes in the network using the Exponential mechanism~\citep{dwork2006our}.
  While the original PrivBayes uses the Laplace mechanism~\citep{dwork2006calibrating} for marginals measurement and directly samples from them, we adjust our implementation to incorporate the Gaussian mechanism and Private-PGM, similar to~\citep{mckenna2019graphical} to unify it with the MST/AIM implementations.
  We refer to this model as PrivBayes+PGM (note: it satisfies ($\epsilon, \delta$)-DP instead of pure DP).
  \item \textbf{MST} first chooses all one-way marginals (corresponding to the columns in the dataset) and then selects a collection of two-way marginals that form a maximum spanning tree (an undirected graph) of the underlying correlation graph.
  \item \textbf{AIM} also starts by selecting all one-way marginals but then chooses higher-order ones iteratively based on their overall importance and comparative error to already chosen ones.
\end{compactitem}

\descr{DP Guarantees.}
Numerous popular DP libraries, including synthetic data generation ones, contain well/long-known DP vulnerabilities such as direct data domain extraction from the input data and floating-point (im)precision~\citep{casacuberta2022widespread, haney2022precision, lokna2023group, annamalai2024you} (see Table~\ref{tab:dp}).
To ensure end-to-end DP, we rely on scientific papers and best practices addressing these issues.

\begin{compactitem}
  \item \textbf{Data Domain.} \dpmm{} allows the data domain (for both numerical and categorical data) to be provided as input.
  Alternatively, we follow \citet{desfontaines2020lowering}'s method to extract min/max for numerical data using half the preprocessing budget $0.5\epsilon_{proc}$ (also implemented in OpenDP).
  \item \textbf{Data Preprocessing.} Since all models operate on discrete data, \dpmm{} has two discretization strategies -- PrivTree~\citep{zhang2016privtree} and uniform.
  The former is a DP tree-based method using $0.5\epsilon_{proc}$ (Laplace mechanism) and is the default since it has shown superior performance~\citep{tao2022benchmarking, ganev2025importance}.
  \item \textbf{Floating-Point Precision.} To defend against floating-point vulnerabilities when sampling from the Gaussian mechanism, \dpmm{} follows \citet{casacuberta2022widespread}'s method, whose effectiveness has been verified by researchers~\citep{desfontaines2023how, lokna2023group} (originally implemented in OpenDP).
\end{compactitem}

\descr{Functionality.}
Next, we discuss a set of features, which are available to all three models in \dpmm{}:

\begin{compactitem}
  \item \textbf{Mixed Data Support:} the training data can contain both numerical and categorical data.
  \item \textbf{Conditional Generation:} generated synthetic data can satisfy any conditions at generation time, without the need to retrain the model or use negative sampling, due to Private-PGM.
  \item \textbf{Public Pretraining:} the model can be pretrained on public data, i.e., extract the data domain, fit data preprocessing, and select the marginals of interest without spending privacy budget.
  \item \textbf{Structural Zeros:} categories or numerical intervals that always have zero counts in the training data can be configured to be preserved in the generated synthetic data.
  Additionally, the accurate modeling of very rare categories or numerical intervals can be suppressed.
  \item \textbf{Max Model Size:} the size of the trained model (in terms of MBs) can be controlled.
  \item \textbf{Serialization:} trained models can be saved and reloaded for later use.
\end{compactitem}

\section{Empirical Evaluation}
\label{sec:emp}
In this section, we evaluate the three marginal models in \dpmm{} (PrivBayes+PGM, MST, and AIM) both in terms of utility and privacy, and compare them to the implementations in other libraries.

\descr{Utility.}
We train models with $\epsilon=1$ (and $\epsilon_{proc}=0.1$ for processing when applicable) using all available implementations on the Wine dataset~\citep{dua2017data}.
We use the default hyperparameters for all implementations (with $\delta=10^{-5}$), while the data domain is assumed to be known and provided as input.
Utility is measured as the average score between real and synthetic datasets across three metrics: marginal (1-way) and pairwise (2-way) similarity, as well as the ability to distinguish real from synthetic records.
All reported scores are averaged across ten trained model instances and ten generated synthetic datasets per trained model.
We visualize the utility and runtime results in Fig.~\ref{fig:comparisson}.

Compared to the original private-pgm, \dpmm{}'s MST and AIM perform on par and even achieve slightly better results, with 1.5\% improvement.
We attribute this to the improved preprocessing strategy, PrivTree, consistent with~\cite{ganev2025importance}.
Moreover, \dpmm{}'s MST runtime is significantly faster due to parallelization when measuring all 1-way marginals.
Unfortunately, AIM takes longer, as smaller marginal estimation errors allow the model to run for more iterations before the budget is spent.
Also, \dpmm{}'s implementations dramatically outperform OpenDP and synthcity by 147\% on average.

\begin{figure*}[t!]
  \vspace{-0.5cm}
	\begin{minipage}{0.49\linewidth}
    \centering
		\includegraphics[width=0.99\linewidth]{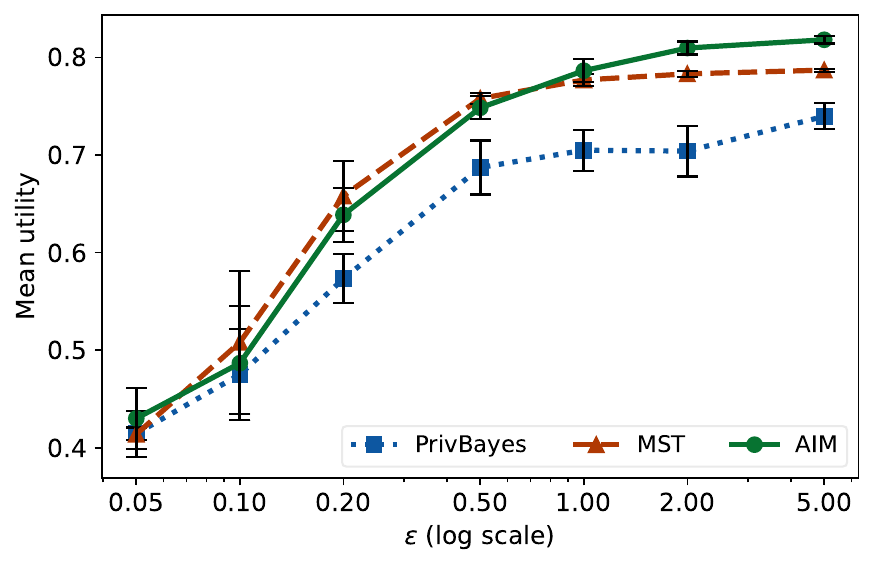}
		\caption{\small Utility-privacy tradeoffs of \dpmm{} on Wine.}
		\label{fig:utility_eps}
    \nextfloat
	\end{minipage}
	\hfill
	\begin{minipage}{0.49\linewidth}
    \begin{subfigure}{0.99\linewidth}
      \centering
      \includegraphics[width=0.85\linewidth]{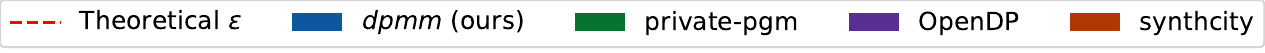}
    \end{subfigure}
    \begin{subfigure}{0.49\linewidth}
      \centering
  	  \includegraphics[width=0.99\linewidth]{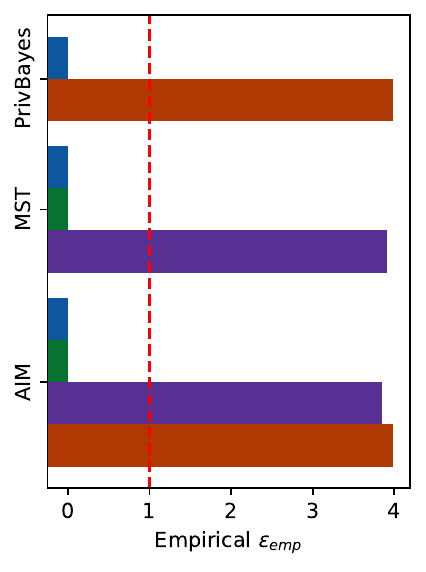}
      \caption{\small AuditSynth~\citep{annamalai2024you}}
  	  \label{fig:overall}
    \end{subfigure}
    \begin{subfigure}{0.49\linewidth}
      \centering
  	  \includegraphics[width=0.99\linewidth]{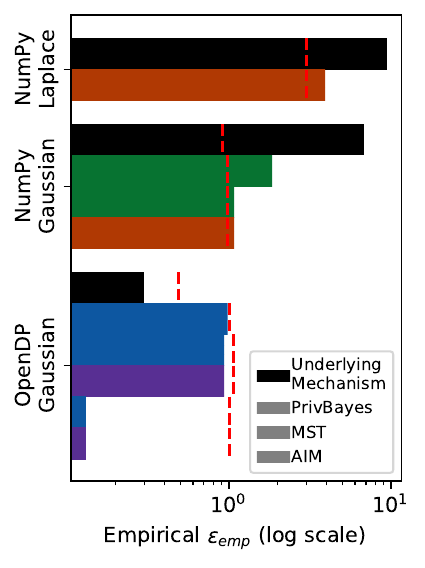}
      \caption{\small Delta-Siege~\citep{lokna2023group}}
  	  \label{fig:float}
    \end{subfigure}
    \caption{\small DP auditing of \dpmm{} and other libraries.}
    \label{fig:audit}
	\end{minipage}
  \vspace{-0.5cm}
\end{figure*}

Additionally, in Fig.~\ref{fig:utility_eps}, we see that all models in \dpmm{} exhibit desirable behaviors, i.e., their utility increases with more available privacy budget.
Consistent with previous research~\citep{tao2022benchmarking, mckenna2022aim, ganev2024graphical}, MST and AIM perform better than PrivBayes+PGM, while AIM edges out MST for $\epsilon>0.5$.

\descr{DP Auditing.}
We also include state-of-the-art DP auditing techniques in \dpmm{}, alongside extensive unit and integration tests, to ensure that our implementations are free from (known) privacy vulnerabilities.
Specifically, we implement two powerful attacks, AuditSynth~\citep{annamalai2024you} and Delta-Siege~\citep{lokna2023group}, which have been used to: i) detect general privacy violations in PrivBayes and MST~\citep{annamalai2024you}, and ii) identify floating-point precision violations in MST and AIM~\citep{lokna2023group}.
To achieve a tight audit (i.e., the empirical privacy loss~$\epsilon_{emp}$ approaches the theoretical guarantee~$\epsilon$) or detect a violation (i.e., $\epsilon_{emp} > \epsilon$), both attacks assume white-box access to the model (the fitted model's parameters and generated synthetic data), a worst-case target record (lying outside the domain of the rest of the data), and worst-case neighboring datasets (constructed from a couple low-dimensional records).

AuditSynth~\citep{annamalai2024you} uses membership inference attacks via a repeated distinguishing game~\citep{nasr2021adversary, nasr2023tight}, in which an adversary attempts to determine whether the target record was part of the model's training data.
We train 1,000 models per implementation ($\epsilon=1$, $\delta=10^{-3}$) and report the resulting $\epsilon_{emp}$s in Fig.~\ref{fig:overall}.
Both synthcity and OpenDP are vulnerable to this class of general attacks as $\epsilon_{emp} >> \epsilon$ (indicating detected privacy violations), primarily because the data domain cannot be passed as input.
In contrast, private-pgm and \dpmm{} exhibit minimal privacy leakage that remains well within expected bounds, assuming the data domain is specified, which is consistent with prior findings~\citep{annamalai2024you, ganev2025importance}.

Delta-Siege~\citep{lokna2023group} relies on the insight that many ($\epsilon$, $\delta$) pairs can be grouped and efficiently audited together, as they result in the same algorithm.
First, we audit the underlying DP mechanisms used by each library (see black bars in Fig.~\ref{fig:float}): NumPy Laplace (synthcity), NumPy Gaussian (private-pgm, synthcity), and OpenDP Gaussian~\citep{casacuberta2022widespread} (\dpmm{}, OpenDP).
Mechanisms relying on NumPy exhibit floating-point precision violations ($\epsilon_{emp} >> \epsilon$), while OpenDP Gaussian does not, as verified by previous work~\citep{desfontaines2023how, lokna2023group}.
Then, we confirm that these trends hold when auditing the full implementations of the three models (see colored bars in Fig.~\ref{fig:float}).
Specifically, we train numerous models ($\epsilon=1$, $\delta=10^{-3}$) and confirm that only \dpmm{} and OpenDP avoid excessive privacy leakage.

\smallskip
Overall, to the best of our knowledge, \dpmm{} is the only library containing state-of-the-art DP synthetic data models and does not exhibit (known) privacy-related vulnerabilities.
While these audits cannot guarantee the absence of all bugs, they provide strong empirical evidence of \dpmm{}’s robustness.

\section{Conclusion}
In this paper, we introduced \dpmm{}, a library for DP synthetic data generation that includes robust implementations of PrivBayes+PGM, MST, and AIM, offering rich functionality.
We hope that \dpmm{} will serve both practitioners and researchers in their adoption and development of DP synthetic data.
We look forward to continuously improving the functionality, trustworthiness, and performance of the library with the support of the community.

\setlength{\bibsep}{2.75pt plus 10ex}
{\footnotesize

\bibliographystyle{plainnat}
}

\appendix

\end{document}